\documentclass[12pt]{article}
\usepackage{epsfig}
\usepackage{latexsym}
\begin{document}
\begin{center}

{ \bf NONLINEAR LASER-INDUCED DEFORMATIONS OF LIQUID-LIQUID
INTERFACES: AN OPTICAL FIBER MODEL}

\vspace{1cm}

Ole Jakob Birkeland\footnote{E-mail:
ole.jakob.birkeland@cggveritas.com} and Iver
Brevik\footnote{E-mail: iver.h.brevik@ntnu.no }

\bigskip

Department of Energy and Process Engineering, Norwegian University
of Science and Technology, N-7491 Trondheim, Norway

\bigskip

\bigskip

Revised version

\bigskip

\end{center}

\begin{abstract}
Experimentally, it turns out that  radiation forces from a
cw-laser on a liquid-liquid interface are able to produce giant
deformations (up to about 100$~\mu$m), if the system is close to
the critical point where the surface tension becomes small. We
present a new model for such a fingerlike deformation, implying
that the system is described as an optical fiber. One reason for
introducing such a model is that the refractive index difference
in  modern experiments, such as those of the Bordeaux group, is
small, of the same order as in practical fibers in optics.  It is
natural therefore, to adopt the hybrid HE$_{11}$ mode, known from
fiber theory, as the fundamental mode  for the liquid system. We
show how the balance between hydrodynamical and radiation forces
leads to a stable equilibrium point for the liquid column. Also,
we calculate the narrowing of the  column radius as the depth
increases. Comparison with experimental results of the Bordeaux
group yields quite satisfactory agreement as regards the column
width.
\end{abstract}


\section{Introduction}

Force exerted by light on continuous matter has attracted
attention since the days of Maxwell. In 1869 he was the first to
predict and calculate the pressure from light reflecting off a
surface;  in 1901 the first attempts of experimental verification
were made by Lebedev, showing that  the light pressure  acts
inwards to a reflecting surface. Poynting \cite{poynting05} later
extended the knowledge of radiation pressure by  considering light
incident from vacuum on the surface of a transparent dielectric,
predicting the irradiation to produce an  outward normal force
irrespective of the angle of incidence. More recently, the
experimental paper of Ashkin and Dziedzic from 1973 is a
pioneering work \cite{ashkin73}. They illumined an air-water
surface by a focused pulsed laser, verifying that the force was
indeed outward directed. Another recent experiment along the same
lines is that of Schroll {\it et al.} \cite{schroll07}.

The radiation pressure problem is related to the well known
Abraham-Minkowski controversy about the correct form of the
electromagnetic energy-momentum tensor in a medium. That question
has been discussed at varying degrees of intensity ever since
Abraham and Minkowski presented their energy-momentum expressions
around 1910. In the simplest case, when the medium is isotropic
and homogeneous, the difference turns up only in the expressions
for the momentum density $\bf g$: in the Minkowski case $\bf
g^M=D\times B$, whereas in the Abraham case, ${\bf g^A=}
(1/c^2){\bf E\times H}$, the latter expression satisfying Planck's
principle of inertia of energy ${\bf g=S}/c^2$, $\bf S$ being the
Poynting vector. For an introduction to the Abraham-Minkowski
problem, the reader may consult M{\o}ller's book on relativity
\cite{moller72}, or the review article \cite{brevik79} of one of
the present authors. There is by now an extensive
literature in this field; some papers are
Refs.~\cite{kentwell87,antoci98,obukhov03,loudon97,garrison04,feigel04,leonhardt06,birkeland07}.
Fortunately, for practical purposes  the difference between the
Abraham and Minkowski predictions goes away in optics because the
influence from the 'Abraham term' fluctuates out. The force can be
calculated from the electromagnetic stress tensor parts only, and
the stress tensors are equal in the two cases. We shall take this
into account in the following, and simply call the  force  the
Abraham-Minkowski (AM) force.

A major reason why radiation forces has attracted increased
interest in recent years is their practical usefulness. In biology
and medicine the need of non-invasive techniques for manipulation
of individual cells or complex molecules has led to the
development of optical tweezers, invented  by Ashkin {\it et al.}
in 1986 \cite{ashkin86}. Furthermore, fluid-interface
instabilities driven by the relatively strong forces from electric
field \cite{taylor69,badie97} represent  the cornerstone of many
industrial processes such as electrospraying \cite{ganan97},
ink-jet printing \cite{oddershede00}, and surface relief printing
\cite{schaffer00}.

An important progress in the application of pressure forces, is to
make use of two liquids in the vicinity of the {\it critical
point}. Then the  surface tension  is significantly diminished,
and the effect of the pressure forces becomes enhanced.
Traditionally, due to the competition with surface forces, the
pressure effect has been rather  minute. For instance, in the
Ashkin-Dziedzic air-water pressure experiment \cite{ashkin73},
with surface tension equal to $\sigma=$73 mJ/m, the deflection of
the water surface acted upon by a pulsed Nd YAG laser was only
about 1 $\mu$m. Working near the critical point, the deflection
can be much higher, about 50 $\mu$m or more. Moreover, in a
two-fluid system of surfactant-coated nano droplets in oil
microemulsions the surface tension can be made more than one
million times smaller than the usual air-water tension (cf, for
instance, Ref.~\cite{rosano87}).

The main purpose of the present paper is to introduce a fiber
model for the large finger-shaped deformations. Interesting work,
both experimental and theoretical, has in recent years been done
in this direction by the Bordeaux group, considering diverse
effects arising from the focusing of laser beams onto a
fluid-fluid interface having extremely low surface tension. Much
of this work, up to 2002, is summarized in the PhD thesis of
Casner \cite{casner02}. There are several recent articles by this
group
\cite{schroll07,casner01,casner01a,casner03,wunenburger06,wunenburger06a}.
There is also a review paper \cite{delville06a}, and related
papers of others such as \cite{hallanger05}. Whereas the first
experimental and theoretical investigations dealt with small
deformations in the hydrodynamic linear regime, the most recent
experiments have demonstrated the appearance of giant surface
deformations of order 50-100 $\mu$m. Large finger-shaped
structures have been observed, as well as liquid jets and shedding
of microdroplets.

A theoretical description of these finger-shaped deformations is
still lacking. Nor will we in the present paper attempt to give a
quantitative estimate of the large deformations. Instead, we will
show one can calculate a stable equilibrium radius of the liquid
column. As is conventional in fiber optics, we take the HE$_{11}$
mode to be dominant for a step-like fiber having a weak refractive
index difference between core and cladding. This dominance occurs
if the refractive index contrast is small, about 0.01. Actually,
this fits quite well with the conditions of the Bordeaux
experiments. It means that the central core becomes exposed to a
laser intensity differing in shape and intensity from the
conventional Gaussian intensity distribution.

There is another advantage of this kind of wave description. The
radiation pressure is predicted to be nonvanishing even on the
vertical walls of the cavity. The magnitude of the pressure is of
the same order as the pressure from a Gaussian beam on a flat
surface at normal incidence.  Such a pressure is not obtainable
using a ray picture of the beam; in such a case, the (orthogonal)
pressure on a parallel wall is simply zero. In our approach this
problem is avoided.

Section 2 gives, for the sake of readability, a brief account of
the theoretical background in the linear case although this
material is strictly speaking not new. In particular, we review
how the equation of force equilibrium is solved when the input
laser intensity is Gaussian. Section 3 then introduces our
step-index fiber model. We use this model, as we indicated, to
show how the balance of hydrodynamical and radiation forces leads
to a stable equilibrium  radius for a given power $P$ in the laser
beam. Comparison with the giant deformations observed in the
Bordeaux experiments shows reasonably good agreement.

\section{Theoretical background, linear theory}

The electromagnetic force density $\bf f$ in an isotropic,
nonconductive and nonmagnetic medium  (see, for instance,
Refs.~\cite{brevik79,stratton41}), is
\begin{equation}
{\bf f}=-\frac{\epsilon_0}{2} E^2 \nabla n^2+
\frac{\epsilon_0}{2}\left[E^2\rho \left(\frac{\partial
n^2}{\partial \rho}\right)_T
\right]+\frac{n^2-1}{c^2}\frac{\partial}{\partial t}{\bf ( E\times
H)}. \label{1}
\end{equation}
We employ SI units so that the relation $\epsilon_0\mu_0=1/c^2$
refers to a vacuum, and let $\epsilon$  be a relative quantity so
that the constitutive relations are ${\bf
D}=\epsilon_0\epsilon{\bf E}$, ${\bf B}=\mu_0 {\bf H}$. The medium
is assumed to be nondispersive. Only the first term in
Eq.~(\ref{1}) contributes in our case; as mentioned above this is
the Abraham-Minkowski (AM) force
\begin{equation}
{\bf f}^{\rm{AM}}= -\frac{\epsilon_0}{2}E^2 \,\nabla{n^2}.
\label{2}
\end{equation}
The geometry is sketched in Fig.~1 (we assume illumination of the
surface from below). The incident wave is taken to be
monochromatic, ${\bf E}^{(i)}={\bf E}^{(i)}({\bf r})e^{-i\omega
t}$. The plane of incidence is formed by by the vectors ${\bf
k}_i$ and $\bf n$; the angle of incidence is $\theta_i$ and the
angle of transmission is $\theta_t$ (see Fig.~1). Let ${\bf
E}_\parallel$ and ${\bf E}_\perp$ be the components of $\bf E$
parallel and perpendicular to the plane of incidence; then the
respective transmission coefficients are
\begin{equation}
T_\parallel=\frac{\sin 2\theta_i\sin
2\theta_t}{\sin^2(\theta_i+\theta_t)\cos^2(\theta_i-\theta_t)},
\label{3a}
\end{equation}
\begin{equation}
T_\perp=\frac{\sin 2\theta_i \sin
2\theta_t}{\sin^2(\theta_i+\theta_t)}. \label{4a}
\end{equation}
Let  $I=\epsilon_0n_1c\langle {E^{(i)}}^2\rangle$ (averaged over
the two polarizations) be the intensity of the incident beam. If
$\alpha$ is the angle between ${\bf E}^{(i)}$ and the plane of
incidence, so that $E_\parallel^{(i)}=E^{(i)}\cos \alpha, \,
E_\perp^{(i)}=E^{(i)}\sin \alpha$, we can write the surface
pressure as \cite{delville06a,hallanger05}
\begin{equation}
 {\bf \Pi}^{AM}=-\frac{I}{2c}\frac{n_2^2-n_1^2}{n_2}\frac{\cos
\theta_i}{\cos
\theta_t}[(\sin^2\theta_i+\cos^2\theta_t)T_\parallel \cos^2\alpha
+T_\perp \sin^2\alpha ]{\bf n}. \label{5a}
\end{equation}
Let now the liquid-liquid interface be represented as
$z=h(r,\theta)$ in cylindrical coordinates ($r=\sqrt{x^2+y^2}$)
and  assume azimuthal symmetry, so that $h_\theta \equiv \partial
h/\partial \theta = 0$. Observing that $\cos
\theta_i=(1+h_r^2)^{-1/2}, \, \sin \theta_i=h_r(1+h_r^2)^{-1/2}$
$(h_r \equiv \partial h/\partial r)$  as well as the corresponding
relations for $\theta_t$ we obtain, upon substitution into
Eq.~(\ref{5a}),
 \begin{equation}
            \Pi = \frac{2n_1I}{c}\frac{1-N}{1+N}f(h_r,\alpha)\,,
        \label{14}
    \end{equation}
    where
    \[N=n_1/n_2<1 \]
    is the refractive index ratio and $f(h_r,\alpha)$ is the
    function
\begin{equation}
        f(h_r,\alpha)= \frac{(1+N)^2}{[N+\sqrt{1+h_r^2(1-N^2)}]^2}\left\{\sin^2\alpha + \frac{1+(3-N^2)h_r^2+
        (2-N^2)h_r^4}{[Nh_r^2+\sqrt{1+h_r^2(1-N^2)}]^2}\cos^2\alpha \right\}.\label{15}
    \end{equation}
Now it turns out that the polarization plays no important part in
the present problem.       This  can be seen by plotting the
normalized radiation force for different values of the angle
$\alpha$ (not shown here; some more details can be found in
Refs.~\cite{hallanger05} and \cite{birkeland08}). Taking $N =
0.986$, a physically realistic value in the Bordeaux experiments,
it is hard even to separate the curves for varying values for
$\alpha$. In the actual experiments the values of $N$  were indeed
near unity. This gives an expression
 for $f(h_r, \alpha) \rightarrow f(N,h_r)$,
averaged over the angle $\alpha$, that is more than accurate
enough for our purpose:
    \begin{equation}
        f(N,h_r)=(1+N^2)\frac{1+(2-N^2)h_r^2+h_r^4+Nh_r^2S}{(N+S)^2(Nh_r^2+S)^2},
        \label{16}
    \end{equation}
where $S\equiv\sqrt{1+h_r^2(1-N^2)}$.

\begin{figure}[htbp]
    \centering
       \includegraphics[height=5.5cm]{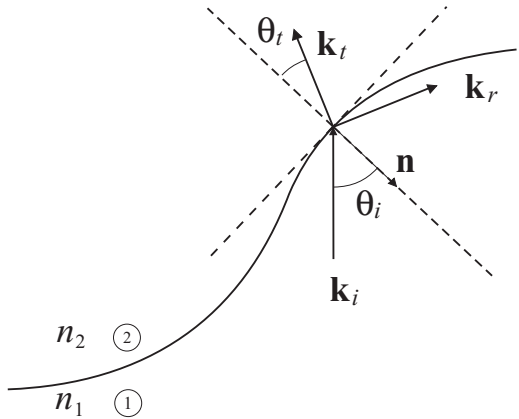}
    \caption{Laser wave incident upon an interface separating media 1 and 2, where the index of refraction $n_1<n_2$.
    The electromagnetic surface force is directed towards the optically thinner medium irrespective of the laser light's direction of propagation.}
    \label{fig:2}
\end{figure}

The system that we shall consider  is comprised of two
near-critical oil-emulsions separated by an interface, caused by
the density contrast between the liquids. It is useful  to
introduce the Bond-number, describing   the relative strength of
buoyancy in comparison to the Laplace force:
    \begin{equation}
        B = \left(\frac{\omega_0}{l_C}\right)^2.\label{4}
    \end{equation}
Here $\omega_0$ is the radius of the beam waist, and $l_C$ is the
capillary length,
    \begin{equation}
        l_C=\sqrt{\frac{\sigma}{(\rho_1-\rho_2)g}},\label{5}
    \end{equation}
where $\sigma$ is the surface tension, $\rho_1$ and $\rho_2$ are
the densities of the lower and upper liquids, and $g$ is the
gravitational acceleration. In the experiments of Casner and
Delville, the Bond-numbers were in the range from $10^{-3}$ to
about 4. If $B\ll 1$, the gravitational   force is  much weaker
than the  surface-tension force.

An advantage of near-critical systems is the possibility to
 tune fluid properties continuously by varying the
temperature. Many  physical quantities scale with temperature as
$\propto[(T-T_{C})/T_{C}]^{\beta}$, where $\beta$ is a constant
(see for instance \cite{freysz85}). Detailed information about
liquid properties near to the critical point is available
elsewhere \cite{casner01a,casner03,delville06a,hallanger05}, and
will not be repeated here.

Consider now the governing equation for the liquid-liquid surface
elevation. Under stationary conditions the elevation is determined
by the balance of radiation pressure, surface tension, and
gravity. Applying Laplace's formula for the pressure difference
$p_2-p_1$, we obtain \cite{delville06a,hallanger05}
 \begin{equation}
        \Delta\rho g\cdot h(r) - \frac{\sigma}{r}\frac{\rm{d}}{\rm{d}r}
\left[\frac{rh_r}{\sqrt{1+h_r^2}}\right] = -\frac{2I(r)}{c}n_1
\frac{n_2-n_1}{n_2+n_1}f(N,h_r)\,,
        \label{22}
    \end{equation}
where $\Delta \rho =\rho_1-\rho_2>0$. The equation can be solved
numerically once $I(r)$ is known.

Let us assume a Gaussian form:
\begin{equation}
I(r)=\frac{2P}{\pi \omega_0^2}\, e^{-2r^2/\omega_0^2}, \label{22a}
\end{equation}
 where $P$ is the total power. It is  convenient to
 introduce the dimensionless variables $R = r/\omega_0, \,  H(R) =
        h(r)/\omega_0$, whereby Eq.~(\ref{22}) can  be written in  dimensionless
        form,
    \begin{equation}
        BH - \frac{1}{R}\frac{\rm{d}}{\rm{d}R}\left[\frac{RH_R}{\sqrt{1+H_R^2}}\right] = -
        F\,{e}^{-2R^2}f(N,H_R),
        \label{24}
    \end{equation}
with
    \begin{equation}
        F   \approx \frac{2P|\partial{n}/\partial\rho|}{\pi c g
         \omega_0\,l_C^2}.\label{25}
    \end{equation}

The function $f(N,H_R)$ is the same as given by Eq.~(\ref{16})
above, except from the  replacement $h_r \rightarrow H_R$. As
before, $B=(\omega_0/l_{C})^2$ is the Bond-number, and $|\partial
n/\partial \rho|=1.22\times10^{-4}$ $\rm m^3/kg$ \cite{casner02}.
The ratio $N$, and the capillary length $l_{C}$, are the only
temperature-dependent parameters.  Thus, there are three
parameters  in Eq.~(\ref{24}), namely the temperature difference
$\Delta T$, the beam waist $\omega_0$, and the beam power $P$.

\begin{figure}[htbp]
    \centering
        \includegraphics[width=1.00\textwidth]{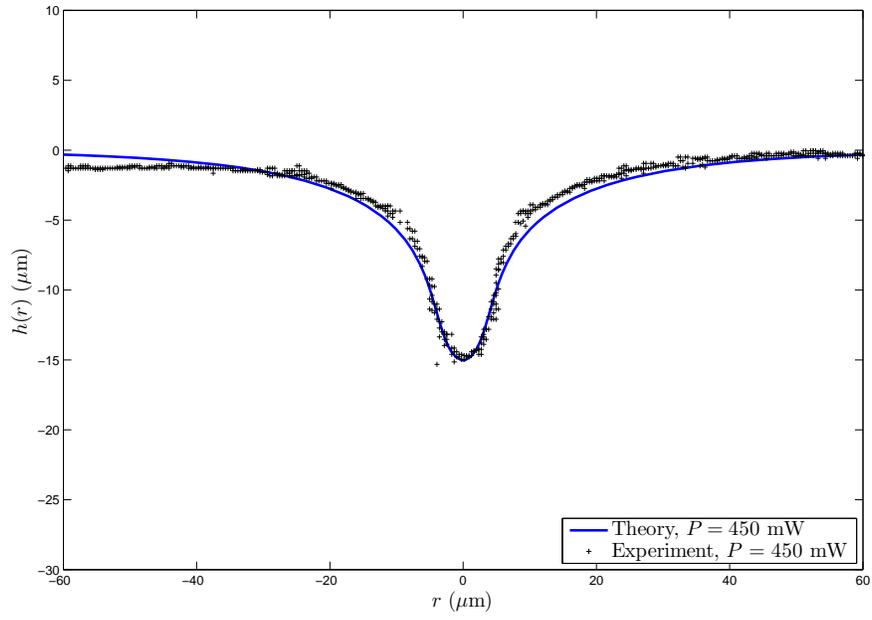}
        \caption{Experimental versus numerical results at temperature such that $\Delta T = T - T_{\rm{C}}=3.5\,$K, and with a laser beam waist of $\omega_0=5.3\,\mu$m at a power of 450$\,$mW.}
    \label{fig:p450dT3expvsteori}
\end{figure}

 Numerical solutions of Eq.~(\ref{24}) are shown
  in Fig.~\ref{fig:p450dT3expvsteori}.
  Similar results are presented  in Ref.~\cite{hallanger05}, except that in the present case
  the theoretical curve is  compared directly with experimental data
  from the Bordeaux group. Profiles of the deformations are extracted
   from images taken by a CCD-camera  (provided  by Jean-Pierre
   Delville, personal communication).
    As is seen, the numerical solution gives excellent
   results for a laser power of P=450 mW, beam waist $\omega_0=5.3\,\mu$m, and $\Delta T = T - T_C=3.5\,$K.
    If the laser power is increased, the correspondence between theory and experiment becomes
    poorer. For instance, insertion of the large power $P=1200$ mW
    in the formulas would lead to a considerable overprediction
    (about 50\%) of the surface displacement.

    This concludes our overview of the linear theory.

\section{Electromagnetic wave analysis}
When the power of the laser is increased (for a fixed beam width)
beyond a certain threshold, the  behavior of the liquid-liquid
interface deformation becomes nonlinear.  It has been observed  that a
"shoulder" appears in the case where  light propagates in the
upward direction (cf., for instance, Ref.~\cite{casner02}). If the
light propagates downwards, the deformations become very large and
cylinderlike, eventually forming a jet. The deformations may even
form a liquid bridge crossing a central layer of fluid. In such a
case  it is natural to suggest that the deformations, or rather
the liquid column, may  be regarded as an {\it optical fiber},
capable of guiding the laser-light due to internal reflections. As
already mentioned, this is the main new idea of the present paper.
Furthermore, the deformations and columns tend to have vertical
walls, which corresponds to $|h_r|\rightarrow\infty$.

Examining the expression (\ref{14}) for the radiation pressure, it
can be seen that (see Fig.~3)
    \[
        \Pi^{\rm{AM}} \rightarrow 0, \quad \rm{when}\quad |h_r|\rightarrow\infty,
    \]
implying that the radiation force vanishes, leaving only the
forces of surface-tension and gravity, which both act against any
deformation of the interface. This result is obviously
unphysical, since both steep-walled deformations, and the liquid
columns, are stationary structures. This means that the following
two usual assumptions: \vspace{0.3cm}

(i) The laser-beam intensity distribution is Gaussian, and stays
Gaussian also in the region being deformed;

(ii) ray optics may be used to describe the path of a light beam
within the deformation, \vspace{0.3cm}

\begin{figure}[htbp]
    \centering
        \includegraphics[height=6cm]{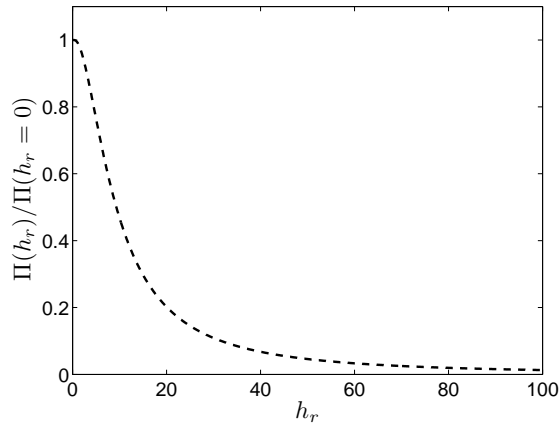}
    \caption{The function $f(N,h_r)$ goes to zero as $h_r$ grows large, i.e. when the walls of the deformation, or the liquid column,
    are (nearly) vertical. The only remaining forces is then the Laplace force and the hydrostatic pressure-force, both acting
    inwards. This implies that the column or deformation should collapse in on itself, which it evidently does not.}
    \label{fig:f(a,h)}
\end{figure}


\noindent must be invalid. Regarding the intensity distribution, one can
argue that the geometrical change of the illuminated interface will
result in new boundary conditions that the electromagnetic field must satisfy.
In turn, this gives rise to a different, and unknown, intensity distribution within
the huge deformations characteristic of the non-linear regime.
Also, ray-optics may give a poor
description of the propagation of the electromagnetic waves within
the structure. In our optical fiber model  the core of radius
$r=a$ is the upper liquid, and the cladding is the surrounding
liquid (the lower liquid). Solving Maxwell's equations for such a
geometry, and relating the power flow through the structure to the
power of the incident laser beam, one should be able to obtain a
more reliable expression for the radiation pressure, and give an
estimate for its magnitude.
\begin{figure}
    \centering
        \includegraphics[width=0.30\textwidth]{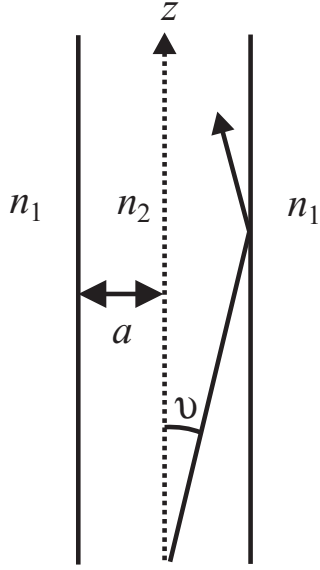}
    \caption{Section of the deformation modelled as an optical step-index fiber. }
    \label{fig:optisk-fiber}
\end{figure}

\subsection{Derivation of the modified radiation force}
Assume as before  plane wave propagation and sinusoidal
time-dependence of the fields, $E,B\propto e^{-i\omega t}$.
Maxwell's equations may be combined to give the following
Helmholtz wave equations for $\bf{H}$ and $\bf{E}$ (cf., for
instance, Ref.~\cite{jackson99}):
    \begin{equation}
        \nabla^2{\bf H} + \frac{n^2\omega^2}{c^2}{\bf{H}} = i\omega\epsilon_0(\nabla n^2)\times \bf{E},
        \label{27}
    \end{equation}
    \begin{equation}
        \nabla^2{\bf{E}} + \frac{n^2\omega^2}{c^2}{\bf{E}} = -\nabla
        \left[ \frac{1}{n^2}(\nabla n^2)\cdot \bf E\right].
        \label{28}
    \end{equation}
 By taking the
$z$-components of Eqs.~(\ref{27}) and (\ref{28}), and eliminating
the transverse field components, one finds the two-dimensional
scalar wave equations for the system,

   \begin{equation}
\nabla_t^2 H_z+\gamma^2H_z-\left(\frac{\omega}{\gamma
c}\right)^2(\nabla_tn^2)\cdot \nabla_tH_z =-\frac{\omega k_z
\epsilon_0}{\gamma^2}{\bf \hat z}\cdot[\nabla_t n^2\times
\nabla_tE_z], \label{29}
\end{equation}
and
    \begin{equation}
        {\nabla_t}^2E_z + \gamma^2H_z - \left(\frac{k_z}{\gamma n}\right)^2(\nabla_{t}n^2)\cdot\nabla_{t}E_z =
         \frac{\omega k_z \mu_0}{\gamma^2n^2}
                                                                                        \hat{\bf{z}}\cdot[\nabla_{t} n^2\times
                                                                                         \nabla_{t}H_z],
        \label{30}
    \end{equation}
where $\nabla_{t}^2=\nabla^2-\partial_z^2$ is the transverse part
of the Laplacian operator. The solutions to these equations are
\begin{equation}
            \{E_z, H_z\}=
  \{A_e, A_h\} J_m(\gamma r)e^{im\theta+ik_zz-i\omega t}, \quad r<a
        \label{31}
    \end{equation}
and
    \begin{equation}
       \{E_z, H_z\}=
\{B_e, B_h\} K_m(\beta r)e^{im\theta+ik_zz-i\omega t}, \quad r>a.
        \label{32}
    \end{equation}
Here, $m$ is an integer, $J_m$ is the $m'$th Bessel function of
the first kind, and $K_m$ is a modified Bessel function. The
radius of the deformation (the core radius) is  $r=a$. The
constants $A_e, A_h, B_e$, and $B_h$ are determined from the
boundary conditions at $r=a$. Furthermore,
  \[
        \gamma^2 = \omega^2n_2^2/c^2-k_z^2 = k_0^2n_2^2-k_z^2 \equiv k_2^2-k_z^2, \quad (r<a)
    \]

    \[
        \beta^2 = k_z^2-\omega^2n_1^2/c^2 = k_z^2-k_0^2n_1^2 \equiv k_z^2-k_1^2,  \quad (r>a)
  \]
are the radial propagation constants.  $k_z$ is the longitudinal
wave number, given by $k_z=k_2\cos\vartheta$, with $\vartheta$
being the angle of propagation. See Fig.~\ref{fig:optisk-fiber}.
Note that $k_0=\omega/c$ refers to the vacuum, $ k_2=k_0n_2$
refers to the core, and $k_1=k_0n_1$ refers to the cladding.

Inserting the solutions (\ref{31})and  (\ref{32}) into the wave
equations (\ref{29}) and (\ref{30}), one can find the radial and
azimuthal field-components. Detailed calculations are given in the
books of Stratton \cite{stratton41} and Okamoto \cite{okamoto06}.
One has to observe that there are multiple solutions to Maxwell's
equations for the step-index geometry, corresponding to different
modes of propagation. We restrict ourselves here to giving the
expressions for the radial and azimuthal components for the
electric field inside the fiber, $r<a$,

    \[        E_r = \left\{ i\frac{k_zA_e}{\gamma}J_m'(\gamma r) - \frac{\omega\mu_0mA_h}{\gamma^2r}J_m(\gamma r) \right\}
                                                                                                e^{im\theta+ik_zz-i\omega
                                                                                                t}\]
            \begin{equation}
            E_{\theta} = -\left\{ \frac{k_zmA_e}{\gamma^2r}J_m(\gamma r) + i\frac{\omega\mu_0A_h}{\gamma}J_m'(\gamma r) \right\}
                                                                                        e^{im\theta+ik_zz-i\omega t}
                                                                                        ,\label{33}
                                                                                        \end{equation}

         \[   E_z = A_eJ_m(\gamma r)e^{im\theta+ik_zz-i\omega
         t},\]

and similarly the $H$-field  for $r<a$,

\[ H_r = \left\{ i\frac{k_zA_h}{\gamma}J_m'(\gamma r) +
\frac{\omega\epsilon_0n_2^2mA_e}{\gamma^2r}J_m(\gamma r) \right\}
            e^{im\theta+ik_zz-i\omega t},\]
            \begin{equation}
H_{\theta} = \left\{- \frac{k_zmA_h}{\gamma^2r}J_m(\gamma r) +
i\frac{\omega\epsilon_0 n_2^2A_e}{\gamma}J_m'(\gamma r) \right\}
e^{im\theta+ik_zz-i\omega t},\label{34}
\end{equation}

\[ H_z = A_hJ_m(\gamma r)e^{im\theta+ik_zz-i\omega
t}. \] The prime on the Bessel functions indicates differentiation
with respect to the argument $\gamma r$.

The aim now is to determine the allowed discrete angles
$\vartheta$ at which the light rays may propagate, corresponding
to the allowed modes of propagation. We begin by defining the
normalized transverse wave numbers as
    \begin{equation}
u \equiv \gamma a = a\sqrt{k_{2}^2-k_z^2}, \label{35}
\end{equation}
\begin{equation}
             w \equiv \beta a =
             a\sqrt{k_z^2-k_1^2}.\label{eq:wavenumbers}\label{36}
    \end{equation}
Furthermore, the wave numbers $u$ and $w$ are related, from
Eq.~(\ref{36}), as
    \begin{equation}
        u^2 + w^2 = k_0^2(n_2^2-n_1^2)a^2 \equiv v^2,
        \label{37}
    \end{equation}
referred to as the \emph{normalized frequency}. The
 boundary conditions are  that the tangential field
components are continuous across $r=a$. From them the general
dispersion relation is constructed, valid for all values of
$N=n_1/n_2 <1$  \cite{okamoto06}:
 \[
    \left[\frac{J_m'(u)}{uJ_m(u)} + \frac{K_m'(w)}{wK_m(w)}\right] \left[\frac{J_m'(u)}{uJ_m(u)}+
        N^2\frac{K_m'(w)}{wK_m(w)}\right]
        \]
        \begin{equation}
        =
m^2\left(\frac{1}{u^2}+\frac{1}{w^2}\right)\left[\frac{1}{u^2}+\frac{N^2}{w^2}\right].\label{38}
    \end{equation}

\subsection{Fundamental mode of the step-index fiber. Calculation of the equilibrium radius}
To simplify the discussion, we shall from now on  consider only
the fundamental mode of the step-index optical fiber. This mode
corresponds to the $m=1$ solution of the scalar wave equation, and
is called a  hybrid mode. The fundamental mode of a step-index
optical fiber is the HE$_{m=1,l=1}$ mode, corresponding to both
$E_z$ and $H_z$ nonzero \cite{okamoto06,snitzer61}, in contrast to
the transverse electric (TE) and transverse magnetic (TM) modes
which correspond to $m=0$  having $E_z$ and $H_z$ equal to zero
respectively, but  which possess low cutoff frequencies.

 The index $l=1$ corresponds to the first root of the dispersion relation
satisfying $k_1 <k_z<k_2$. The HE$_{11}$ mode has no cutoff
frequency; this is the reason why it is regarded as  the
fundamental mode of a step-index fiber. For a typical fiber in
optics, the relative refractive index difference $\Delta
=1-n_1/n_2$ is of the order of 0.01. This suits well with the
conditions of the Bordeaux experiments: with the typical value of
$N=0.986$ as mentioned above, we get $\Delta=0.014$, which is
quite appropriate. Under such conditions the confinement of light
in the core is not so tight (this is called the weakly guiding
approximation). The approximation $N \approx 1$ allows one to
simplify the theory of optical fibers very much, and to obtain
clear-cut results.  The accuracy of adopting the  HE$_{11}$ mode
as a basis in our model should be more than sufficient.

 A sketch of the intensity distribution for this mode is given in
Fig.~5.  Numerically solving the dispersion relation (\ref{38})
for a given fiber radius $a$, we obtain the wave number $k_z$
corresponding to that particular radius. Repeating this procedure
for a range of radii, and with all field components being known,
we can  calculate the components of the Abraham-Minkowski surface
force density on the interface of a fiber with a given diameter.
The expression for the AM surface pressure reads,

    \begin{equation}
        {\bf \Pi}^{AM}=\frac{\epsilon_0}{2}(n_2^2-n_1^2)\left[E_{\theta}^2+{E}_z^2
        + N^{-2}E_r^2\right]_{r=a-}{\bf n}.
        \label{39}
    \end{equation}
    Here
    \begin{equation}
    {\bf n}=(1+h_x^2+h_y^2)^{-1/2}(h_x, h_y, -1)
    \label{40}
    \end{equation}
    is the normal vector to the interface, pointing  from the
    internal medium 2 to the external medium 1.
    If ${\bf S}=\frac{1}{2}(\bf E\times H^*)$ denotes the Poynting vector,
 the total power $P$ carried by the optical fiber plus the
    cladding is given by
    \begin{equation}
        P = \int_0^{2\pi}\int_0^{\infty}S_zr\,{d}r\,{d}\theta =
                                                \frac{1}{2}\int_0^{2\pi}\int_0^{\infty}(E_rH_{\theta}^*-E_{\theta}H_r^*)r\,{d}r\,{d}\theta.
        \label{41}
    \end{equation}
Some calculation yields
    \begin{equation}
        P = P_{{core}} + P_{{clad}} = \frac{\pi}{4}\epsilon_0ca^2|A|^2\left[n_2F(J_0,J_1) + n_1\frac{J_1^2}{K_1^2}G(K_0,K_1)\right],
        \label{42}
    \end{equation}
where
    \begin{equation}
        F(J_0,J_1) \equiv \frac{k_z^2a^2}{u^2}\left[(1+s^2)(J_0^2+J_1^2)-\frac{2}{u^2}(1+s)^2J_1^2\right] + J_0^2 + J_1^2 - \frac{2J_0J_1}{u},
        \label{43}
    \end{equation}
and
    \begin{equation}
        G(K_0,K_1) \equiv \frac{k_z^2a^2}{w^2}\left[(1+s^2)(K_1^2-K_0^2)+\frac{2}{w^2}(1+s)^2K_1^2\right] + \frac{2}{w}K_0K_1.
        \label{44}
    \end{equation}
Here, it is understood that $J_n, K_n = J_n(u),K_n(u)$. The
constant $A$ is equal to $A_{{e}}$ above, and $B_{{e}}$ has been
eliminated via the boundary condition $E_z(a-)=E_z(a+)$. The
parameter $s$ is given by
    \begin{equation}
      s \equiv \frac{\left(\frac{1}{u^2}+\frac{1}{w^2}\right)}{\left[\frac{J_1'(u)}{uJ_1(u)} + \frac{K_1'(w)}{wK_1(w)}\right]} \approx -1
        \label{eq:sdefinisjon}
    \end{equation}
(cf. Ref.~\cite{okamoto06}). When the steepness is large
($h_r\gg1$) the expression for the Laplace force in the radial
direction takes on the simple form
\begin{equation}
    f_L = -\lim_{h_r\rightarrow\infty} \, \frac{\sigma}{r}\frac{d}{d r}\frac{rh_r}{\sqrt{1+h_r^2}} = -\frac{\sigma}{r}.
    \label{eq:laplacesteep}
\end{equation}
The force balance may then be written as
\begin{equation}
    -\Delta \rho g h(r=a) +\frac{\sigma}{a} = \Pi^{AM}(r=a),
    \label{eq:kraftbalanse}
\end{equation}
This equation can  be interpreted physically: The right hand side
is the AM surface force density acting outwards. It is positive
because $n_2>n_1$. This force is balanced by the Laplace force
$\sigma/a$ acting inwards, plus the net hydrostatic pressure
$-\Delta\rho gh(r=a)=-(\rho_1-\rho_2)gh(r=a)$ which also acts
inwards  because $\rho_1>\rho_2$. Notice that $h(r)$ is a negative
quantity. We do not regard $a$ as a fixed parameter, but rather
determine the magnitude of the radiation pressure for
 a range of given values of $a$, and subsequently find the corresponding
equilibrium radii where the radiation pressure exactly balances the forces
 of surface tension and hydrostatic pressure. Ergo, the thickness of the
liquid column will vary with depth. For a given depth, the hydrostatic pressure
 difference is known, and then it is easy to determine the new equilibrium
 radius $a$. Notice that we still regard the structure as an ideal step-index
fiber, although in fact it is not. The fiber rather resembles a tapered cylinder,
which would require a different solution to Maxwell's equations to get exact
 results. But in our  model,  it seems reasonable to regard the fiber locally
(with depth) as an optical cylindrical guide with zero taper.
\begin{figure}
    \centering
        \includegraphics[width=1.00\textwidth]{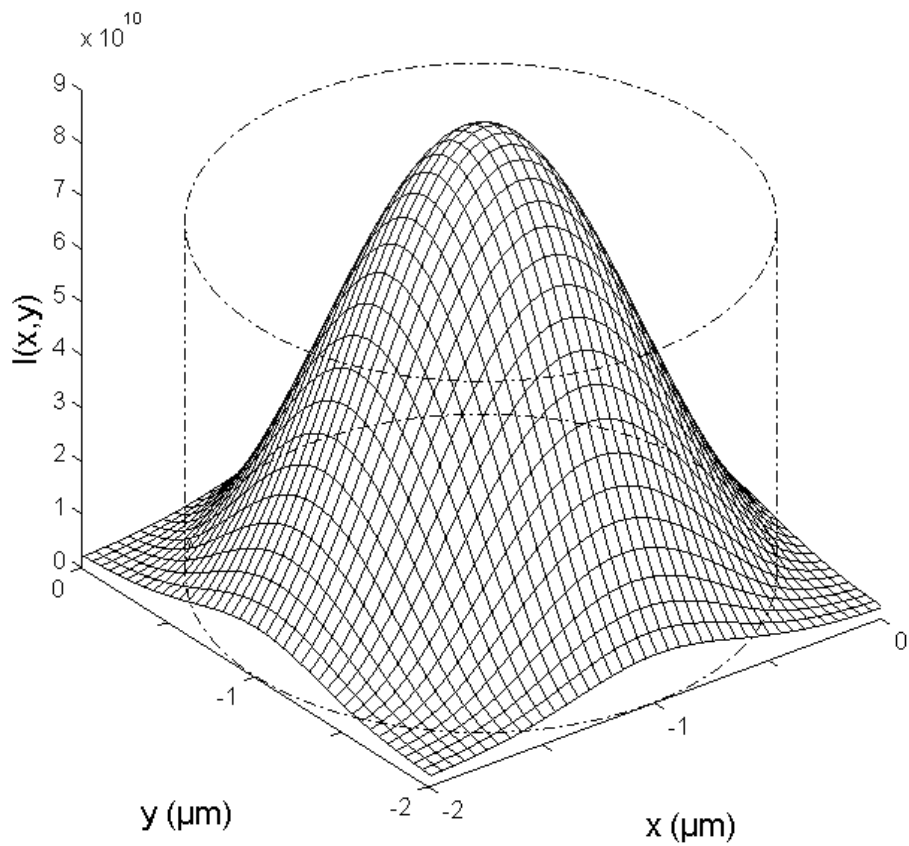}
    \caption{The intensity distribution for the HE$_{11}$-mode in a step-index fiber of radius $a=2\,\mu$m, at a laser beam power
    of 500 mW, and with $\Delta T=3.5$\,K. Broken lines indicate the fiber boundary.}
    \label{fig:HE11intensitet}
\end{figure}

%

\vspace{0.1cm}
 Figure 6 shows how the electromagnetic radiation AM
force varies versus the fiber radius $a$ , for given values of
$\Delta T$ and $P$. The combined pressure of surface tension (the
Laplace force) and gravity (hydrostatic pressure) is also plotted.
Intersect points correspond to equilibrium radii. Notice the
middle set of curves with power $P=150\,$mW. Here, the termination
point of the liquid column can clearly be seen. As the depth
increases beyond $550\,\mu$m, the outward acting AM force can no
longer support the liquid column due to increasing hydrostatic
pressure and surface tension. The model therefore predicts that at
these parameter values, the liquid column can  maximally reach a
depth of $\sim 500\,\mu$m before collapsing, in good
correspondence with  observed break-ups of the liquid columns into
droplet spraying jets at depths of around 1 $\mu$m. In our simple
model we do not take into account the  fluid flow inside the
column; nevertheless, the model offers a simple and intuitive
explanation of the observed pinch off of the columns.

 With increasing temperature difference $\Delta T$ the radiation
force becomes weaker relative to the Laplace force and the
hydrostatic pressure  (details not shown here). This behavior is
expected, since the surface tension is temperature dependent and
an increase in $\Delta T$ leads to a higher surface tension. Now
the hydrostatic pressure  slightly increases as the density
contrast increases, but this is not of great importance since the
term $\Delta \rho gh$ is very small. The net behavior is that for
higher $\Delta T$ the equilibrium radii become smaller.

\vspace{0.2cm}

In Fig.~7, the two left panels refer to temperature
differences, $\Delta T=\{3.5, 1.5\}$ K, and to a moderate power,
$P=$200 mW. These  parameter values make our model quite
justifiable, as can be seen from  the following argument:

\begin{itemize}

\item  When $\Delta T$ is small the relative refractive index
becomes also small, in our case amounting to 0.55\% (i.e.,
$\Delta=1-n_1/n_2=0.0055)$. The microemulsion liquid used in the
Bordeaux experiments had a temperature dependent refractive index
of around 1.46 (\cite{casner02}).
 The laser used was an argon laser operating in the
TEM$_{00}$ mode, with a vacuum wavelength of $\lambda_0=0.514 \,
\mu$m. From these values we calculate the normalized frequency
\begin{equation}
v=k_0a\sqrt{n_2^2-n_1^2}=1.88 a\, [\mu\rm{m}]. \label{36}
\end{equation}

\item  The second point is that for moderate powers the
equilibrium radii for the liquid column become small. Physically,
the reason is that the AM outward pressure is not strong enough to
widen the column very much. The two left panels in Fig. 7 show
that for $P=200$ mW the radii are about 1 $\mu$m. It is
instructive here to compare with Fig.~3 in Okamoto's book
\cite{okamoto06}, which shows the occurrences of multiple modes in
a typical single-step fiber. As long as $v$ is less than about
2.4, only the basic HE$_{11}$ mode appears. For increasing $v$, up
to  3.5, perhaps even higher, there occurs a mixture with the next
HE$_{21}$ mode, but we find it reasonable to expect that the basic
mode is the dominant one at least up to about $v=3$. Insertion of
$a=1$ $\mu$m in Eq.~(\ref{36}) yields $v\approx 2$. Accordingly,
the single-mode model ought in this case to be a viable one. Also,
we note from Fig.~7 that the {\it uniform-cylinder} model of the
column turns out to be a good approximation, as the slope of the
column is large, $|h_r| \sim 10^4$.
\end{itemize}

Actually, it seems that the above limits for the applicability of
the single-mode model may be stretched considerably. The third
panel of Fig.~7 shows the predicted column width when the power is
high, $P=1750\, $mW. This is in violation of the single-mode
condition, but still the calculated results yield a figure
visually quite similar to   the experimental deformation shown in
the rightmost frame. The last-mentioned frame, taken from
Ref.~\cite{casner02}, shows a long jet with large radius,
$a\approx 7 \, \mu$m.  The good agreement between theory and
experiment indicates that the single-step model is quite robust.

\vspace{0.2cm}

The last point that we shall focus attention on, is the observed
termination, or "pinch off", of the liquid column at large depths
where the radius becomes small. We can obtain a semi-quantitative
explanation of this effect on the basis of our model. Namely, when
$a$ becomes smaller than about 0.6 $\mu$m the repulsive AM
radiation force is no longer able to withstand the inward directed
Laplace and hydrostatic forces (there is no equilibrium point in
Fig.~6), and the column has to collapse. What one actually
observes in experiments, is that the column breaks up into a row of small
droplets. In Fig.~7 (a) and (b), the theoretical pinch-off of the columns is
seen to occur at depths of around $1000\,\mu$m.
At equal power, the column stretches deeper when
the surface tension is lower ($\Delta$ T =1.5$\,$K).
It is moreover observed that the break-up occurs in the
case of illumination from above only. Our model is however unable
to explain that particular effect. No theory for it seems to be
known.

\begin{figure}
    \centering
        \includegraphics[width=1.00\textwidth]{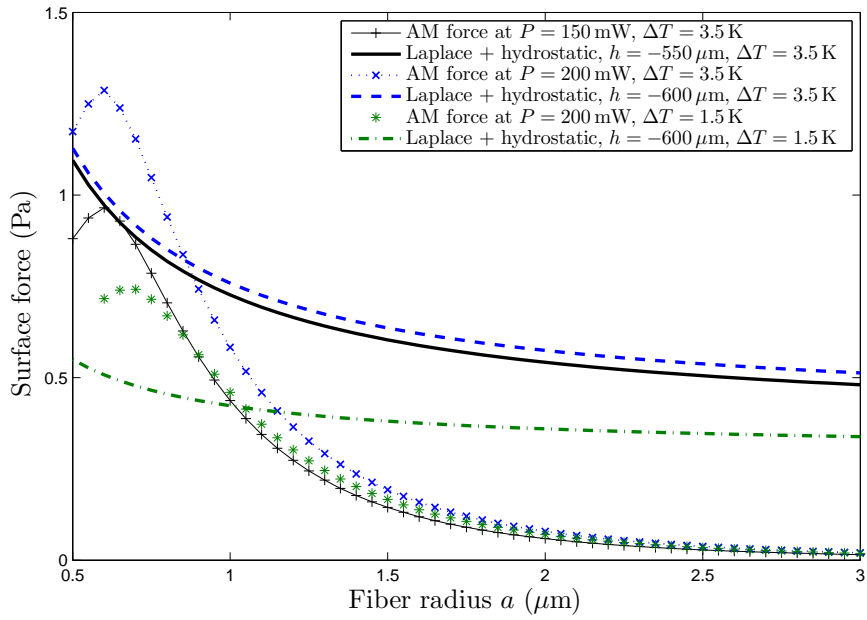}
    \caption{Color online. The electromagnetic radiation pressure (AM force) plotted as a function of varying fiber radius $a$, at
     given temperature $\Delta T$ and beam power $P$.}
    \label{fig:FAMvsFLaplaceHE}
\end{figure}

\newpage

\begin{figure}
    \centering
        \includegraphics[width=1.00\textwidth]{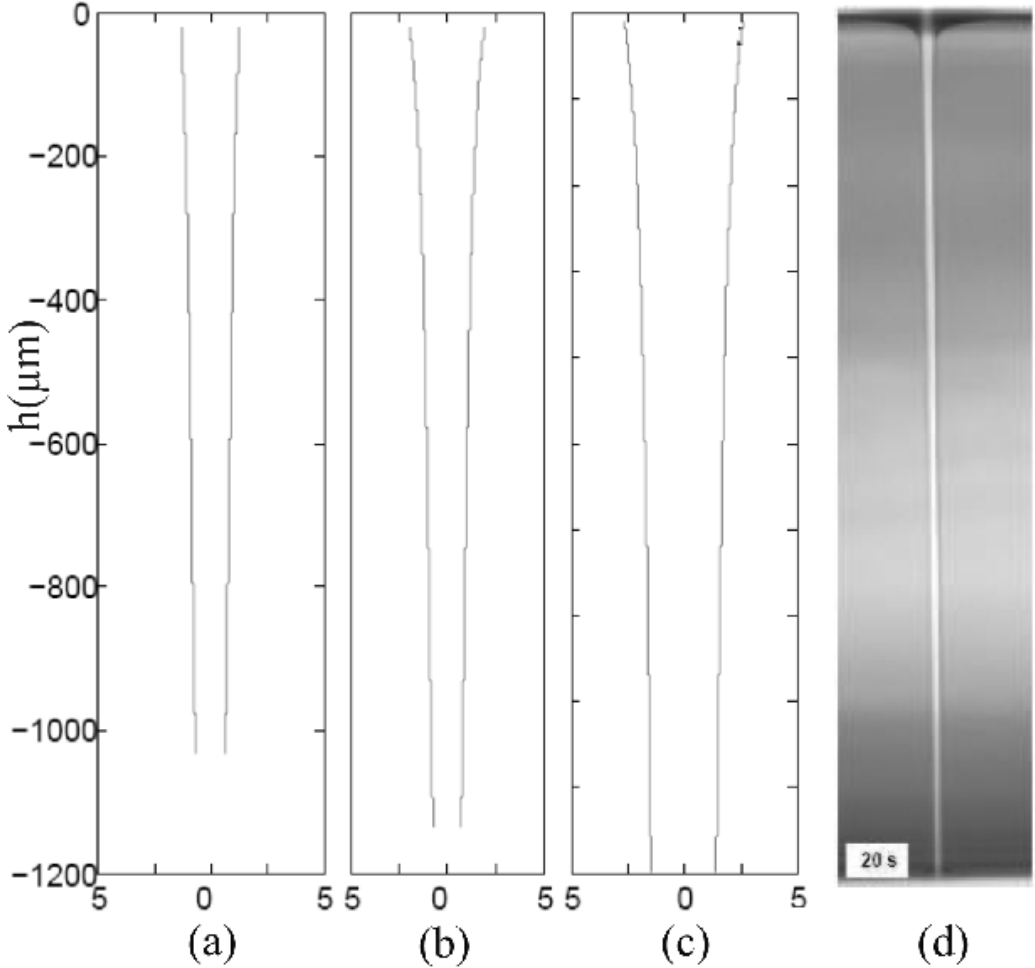}
    \caption{Color online. Calculated interface deformation resulting from the balance
     between the modified AM-force, the Laplace force,
    and the hydrostatic  force. Panel (a) and (b) have $P=200$, and $\Delta T=\{3.5, 1.5 \}$K, respectively. At this moderate
    power level and values of $\Delta T$, the
radii of the columns are, according to the model, around 1 $\mu$m,
thus satisfying the single-mode condition of the step-index fiber.
At high power $P=1750\,$mW and $\Delta T=6\,$K, shown in panel
(c), the  column is predicted to be slightly wider with a radius
from 1.5 to 3 $\mu$m. The picture (d) is taken from
Ref.~\cite{casner02}, and shows the liquid column resulting from
illuminating the surface from above with a laser at $P=1750\,$mW,
and with $T-T_C=6\,$K. In Ref.~\cite{casner02} it is stated that
the jet is about 1000$\,\mu$m deep, with a radius around $7\,
\mu$m, i.e., of the same order of magnitude as that calculated in
our model.}
    \label{fig:kolonnefest}
\end{figure}

\newpage

\section{Summary }

Let us summarize our work as follows: \vspace{0.2cm}

(1) Our main idea has been to model the large fingerlike
deformations, the liquid columns and liquid jets seen in laser
optics experiments, as optical waveguides or fibers. We have
identified the electric field components of a step-index fiber,
identifying the HE$_{11}$ mode as the fundamental mode. The
HE$_{11}$ mode is a hybrid mode, thus more complicated than the
conventional TE and TM modes in that the axial electromagnetic
fields $E_z$ and $H_z$ are not zero, but it is precisely this mode
that is the dominant one in conventional step-index fibers when
the index contrast $N=n_1/n_2$ is close to 1 \cite{okamoto06}.
 The fact that the giant deformations
are seen experimentally  in the vicinity of the critical point
when $N\approx 1$, makes our fiber model very natural from a
physical viewpoint. Moreover, the experiments performed by the
Bordeaux group
\cite{casner02,schroll07,casner01,casner01a,casner03,wunenburger06,wunenburger06a}
correspond just to $N\approx 1$.

(2) In Sect.~2 we reviewed the linear theory, showing how the
elevation of the interface can be calculated with quite good
accuracy for low laser powers $P$ when comparing with the Bordeaux
experiments. No attempt was however made to calculate the
elevations in the nonlinear, high-power case. In particular, we
did not consider the formation of a "shoulder" as seen
experimentally upon illumination from below. No theory for the
nonlinear case seems so far to exist. Instead, we made use of our
model to show how the balance between hydrodynamical and radiation
forces leads to the establishment of stable radii for the liquid
column. See Fig.~6. This means that a physical explanation is
given for the establishment of these giant structures. We
emphasize that such a result is not obtainable from conventional
radiation theory assuming a Gaussian intensity profile for the
incident laser beam. Our intensity distribution described by the
HE$_{11}$ mode is different from the Gaussian form. See the more
detailed discussion on this point in the first part of Sect.~3.

(3) Our comparisons with experiments are all made with reference
to  the Bordeaux group. Figure 7 shows how our calculated values
for the column radius $a$ correspond quite well with the values
observed. In particular, the narrowing of the width for increasing
depths is reproduced. Such a narrowing, by a factor of 2 or more,
is typically seen in experiments (Jean-Pierre Delville, personal
communication).

\section*{Acknowledgment}

We thank Jean-Pierre Delville for valuable information about the
recent Bordeaux experiments.

\vspace{2cm}

{\bf Note added}

\bigskip
After the above work was completed, we have become aware of two
recent papers of the Bordeaux group
\cite{brasselet08,brasselet08a}. These papers report on both
theoretical and experimental work. As for the theoretical parts,
they are  closely related to our approach above.


\end{document}